\documentclass{ws-ijmpe}
\usepackage{graphicx}

\begin{document}

\newcommand{\dphi}{$\Delta\phi$ }
\newcommand{\mom}{$p_T$ }
\newcommand{\gevc}{GeV/$c$ }
\newcommand{\flow}{$v_2$ }

\markboth{Anne Sickles}{Jet Correlations with Identified Particles
from PHENIX}

\catchline{}{}{}{}{}

\title{Jet Correlations with Identified Particles from PHENIX:
Methods and Results}

\author{Anne Sickles for the PHENIX Collaboration}

\address{Brookhaven National Laboratory, Building 510C\\
Upton, NY 11973,
United States
\\
anne@bnl.gov}

\maketitle

\begin{history}
\end{history}

\begin{abstract}
Azimuthal angle two particle correlations have been shown
to be a powerful probe for extracting novel features of the
interaction between hard scattered partons and the medium
produced in Au+Au collisions at RHIC.  At intermediate $p_T$, 2-5GeV/c,
the jets have been shown to be significantly modified in both their
particle composition and their angular distribution compared to p+p
collisions.  Additionally, angular two particle correlations
with identified hadrons provide information on the possible role of
modified hadronization scenarios such as partonic recombination, which
might allow medium modified jet fragmentation by connecting hard
scattered partons to low $p_T$ thermal partons.

PHENIX has excellent particle identification capabilities and
has developed robust techniques for extracting jet
correlations from the large underlying event.  We present
recent PHENIX results from Au+Au collisions for a variety of $p_T$
and particle type combinations. We also present p+p measurements
as a baseline. We show evidence that
protons and anti-protons in the $p_T$ region of enhanced baryon
and anti-baryon single particle
production are produced in close angle pairs of opposite charge and that
the strong modifications to the away side shape observed for charged
hadron correlations are also present when baryons are correlated.
\end{abstract}

\section{Introduction}
One of the most surprising results from the Relativistic Heavy Ion
Collider (RHIC) has been the large increase in the $p/\pi^+$ and
$\bar{p}/\pi^-$ ratios at intermediate $p_T$, 2--5\gevc~\cite{ppg015}.
Studies of  $\Phi$ mesons~\cite{ppg016} and $\Lambda$ baryons~\cite{starlambda}
have indicated that the origin of the excess is related to
the number of valence quarks rather than particle mass.  
Baryon and meson differences have also be studied by measuring
the elliptic flow, $v_2$, of identified particles which have also been shown
to scale with the number of valence quarks~\cite{ppg022,ppg062,starqn}.
In this same \mom range in p+p collisions particle production
shifts from soft, non-perturbative processes to hard parton-parton
scattering followed by jet fragmentation~\cite{starxt}.
The valence quark dependence of these effects has inspired a class
of models based quark recombination; hadronization is modeled
not by fragmentation, but by quarks close together in phase
space coming together to form hadrons
In some of these models intermediate \mom hadrons
primarily come from soft quarks~\cite{friesprc}.  In other
models quarks from jet fragmentation are allowed to recombine 
with soft quarks~\cite{hwa1}. 
All models extend the \mom range of soft physics and start
with thermalized quark degrees of freedom.

Two particle correlations have been used to determine whether
the baryon excess is associated with hard or soft processes and
to explore in detail baryon/meson differences at low and intermediate
$p_T$.  A systematic study of these correlations will  
allow discrimination between different hadronization scenarios
and measurement of the role of hard scattering at intermediate $p_T$.
Here we present a selection of the recent PHENIX results
from these correlations.

\section{Experimental Details}
\subsection{Two Particle Correlations}
Correlations are measured between two classes
of particles, {\it triggers}
and {\it associated particles}.  
The data presented here are from the 2004 Au+Au $\sqrt{s_{NN}}$=200~GeV RHIC run.
All particles are charged tracks reconstructed
in the PHENIX drift chambers.  Particle identification is done via time of flight.
The start time is provided by the PHENIX Beam-Beam Counters and the stop 
time is measured by either the high resolution time of flight or
the lead-scintillator electromagnetic calorimeter, which provide $K/p$
separation to $\approx$4.0~\gevc and $\approx$2.5~\gevc, respectively.

The azimuthal angular difference between 
the trigger and associated particle, \dphi, is measured.  
The non-uniform \dphi acceptance is corrected for with mixed
pairs where the two particles are from different events.
The associated particle reconstruction efficiency is corrected for
by matching the observed single particle spectra to those measured
in Ref.~\cite{ppg026}.
Correlations from elliptic flow are 
removed by using \flow values measured separately~\cite{ppg072}.  
Remaining yield
is attributed jet correlations, $J(\Delta\phi)$.  
Acceptance corrected \dphi distributions are then described by:
\begin{equation}
\frac{1}{N_{trig}}\frac{dN}{d\Delta\phi} = 
B(1 + 2 v_2^{trig} v_2^{assoc} \cos (2\Delta\phi)) + J(\Delta\phi)
\label{eqflow}
\end{equation}
where $B$ is the combinatorial background level and $v_2^{trig}$ and
$v_2^{assoc}$ are \flow values for triggers and associated particles,
respectively.  $N_{trig}$ is the total number of triggers observed. 
The determination of $B$ is discussed in Sect.~\ref{back_sect}.
In order to quantify the centrality dependence of the 
jet correlations $J(\Delta\phi)$ is integrated over \dphi
within which particles are expected to be from
the fragmentation of the same jet, {\it near-side},
or opposing di-jet, {\it away-side}.  These integrated values
are the average conditional yield of associated particles per trigger.

\subsection{Combinatoric Background Subtraction Procedures}
\label{back_sect}
In two particle correlations a small fraction of
the pairs come from the jet-like source which is to
be measured.  The rest of the pairs come from
other sources in the event, pairs where each particle is
from a different jet, one particle is from a jet and
the other is not, or where both particles are from soft
processes; these pairs are called the combinatorial background.
Unfortunately, the combinatorial background grows faster
than the jet-like signal so extraction of the jet-like 
signal becomes more sensitive to the background normalization
in central collisions. 

PHENIX uses two methods to determine $B$ in the correlations
presented here. 
The first, the absolute subtraction method, is
described in detail below.  This method has
the advantage that it requires no assumption about
the shape of the jet correlations in \dphi. The second method, 
zero yield at minimum (ZYAM),
makes the assumption that there is a region in $\Delta\phi$ where there is
the jet yield is zero.  It is described in detail elsewhere~\cite{ppg032}.

The absolute subtraction method uses a convolution of the
single particle rates to determine the combinatorial pair
rate.  The total number of combinatorial pairs in
the event sample, under the assumption that the jet signal is the
only source of correlated pairs, is:
\begin{equation}
N_{comb} = \langle n_{trig} \rangle \langle n_{assoc} \rangle N_{events}
\label{eqcomb}
\end{equation}
where $\langle n_{trig} \rangle$ and $\langle n_{assoc} \rangle$
are the average number of triggers and partners per event
and $N_{events}$ are the total number
of events.  Normalizing $N_{comb}$ by the total
number of triggers as in Eqn.~\ref{eqflow} gives:
\begin{eqnarray}
\frac{\langle n_{trig} \rangle \langle n_{assoc} \rangle N_{events}}{N_{tri
ggers}} &
= & \frac{N_{trig} \langle n_{assoc} \rangle}{N_{trig}}\\
 & = & \langle n_{assoc} \rangle
\end{eqnarray}
Thus, the combinatorial background normalization 
is simply:
\begin{eqnarray}
 \int_0^{\pi} B d\Delta\phi  &=  &\langle n_{assoc} \rangle \\
B & = & \frac{\langle n_{assoc}\rangle}{\pi}
\label{ncombeq}
\end{eqnarray}

However, the assumption made prior to Eqn.~\ref{eqcomb}
is not completely valid, there are an additional correlations due to
fluctuations of the particle multiplicity.
The more central events within a bin have, on
average, a higher 
number of pairs
than those from the lower centrality part of the bin.
Thus, the combinatorial background level 
is higher because it is biased toward higher multiplicity events.
These correlations increase as the relative width of the trigger
and associated particle multiplicity distributions increase.
 To minimize these effect  the analysis is performed
separately in  fine centrality 
bins, 5\%. This value is near the resolution of the centrality
determination.
The final results in wide centrality bins are the
 average of the fine centrality binning  weighted by the
number of triggers in each bin.


The magnitude of the remaining multiplicity correlations  is estimated by parameterizing
$\langle n_{trig} \rangle$ and $\langle n_{assoc} \rangle$ as a function of a 
centrality parameter, the number of participating nucleons 
 ($N_{part}$) or the number of binary collision ($N_{coll}$).  
Monte Carlo events are generated with $N_{part}$ and $N_{coll}$ 
distributions taken from a Glauber model~\cite{glauber}.
Trigger and associated particle multiplicities are taken to be 
distributed according to a Poisson distribution
with a mean given from the parameterization.   
These Monte Carlo events contain pairs whose only correlation is due to
the $N_{part}$ ($N_{coll}$) of the event.   The multiplicity
correlations are quantified by  a parameter, $\xi$, defined as the ratio
of the observed pair rate to the combinatorial background level 
as calculated under the assumption of no
multiplicity correlations  (Eqn.~\ref{eqcomb}).  
The systematic error on this
procedure comes from varying both the multiplicity parameter 
between $N_{part}$ and $N_{coll}$ and  the
parameterizations of $\langle n_{trig} \rangle$ and $\langle n_{assoc} \rangle$. 
The combinatorial background level $B$ in Eqn.~\ref{ncombeq} 
is increased by a factor of  $\xi$.
 The value of $\xi$ depends on the shape of $\langle n_{trig} \rangle (N)$
and $\langle n_{assoc} \rangle (N)$; a strong
 dependence of the multiplicity on centrality leads to
 larger values of $\xi$ because there is
more of a difference between the central and peripheral 
edges of the centrality bin.
In central, 0-5\%, collisions $\xi\approx$1.002 and in peripheral, 60-65\%,
collisions $\xi\approx$1.2.


\section{Results and Discussion}

\subsection{Identified Trigger Correlations}
\label{sectidtrig}

\begin{figure}[t]
\includegraphics[width=\textwidth]{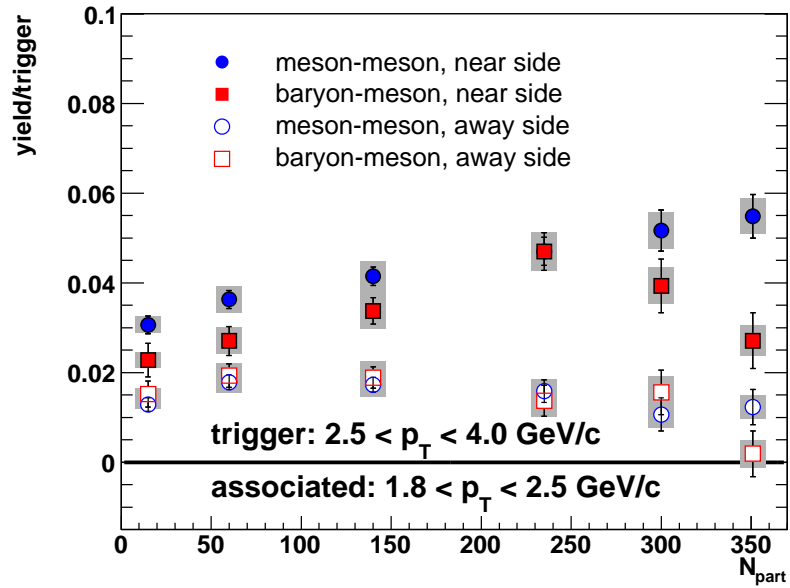}
\caption{Yield per trigger on the near, $\Delta\phi<0.94$rad (solid points) and away 
$\Delta\phi>$2.2rad (hollow
points) side for baryon-meson (squares) and meson-meson (circles) 
correlations as a function of $N_{part}$.  Triggers have 2.5$<p_T<$4.0~\gevc
and associated particles have 1.8$<p_T<$2.5~GeV/$c$.
Error bars are statistical
errors and the shaded boxes show the systematic errors.  There is a
13.6\% normalization error which moves all points together.} 
\label{fig7}
\end{figure}

Fig.~\ref{fig7} shows
near side meson-meson correlations (filled circles) as a function of the number
of participating nucleons, $N_{part}$~\cite{ppg072}.
The background has been subtracted by the absolute subtraction
method.
The yield per trigger rises linearly with increasing
$N_{part}$.  The baryon-meson yield per trigger (filled squares)
also rises linearly for $N_{part}<$250.  In
more central collisions the yield per trigger
decreases and in the most central collisions is
consistent with the peripheral value.  The
agreement between baryon-meson and meson-meson
centrality dependence at $N_{part}<$250 is consistent
with a picture where both trigger types come primarily
from the same source and are associated with an
increasing number of associated particles.
The difference between the baryon and meson triggers
for $N_{part}>$250 could indicate baryon production at
high centralities is dominated by a different source.
However, two particle correlations can only measure the
average number of associated particles per trigger,
not the fraction of triggers which have associated particles.

This ambiguity can be addressed by measuring away-side, 
$\Delta\phi>2.2$rad, yields as a function of trigger
particle type.
Since the partons that become the near and away
side jets are moving away from each other
their fragmentation should be independent.  Fig.~\ref{fig7} also
shows the away side meson yields for baryon and meson triggers
(hollow points).
No significant difference is seen between the trigger types.

\begin{figure}
\includegraphics[width=\textwidth]{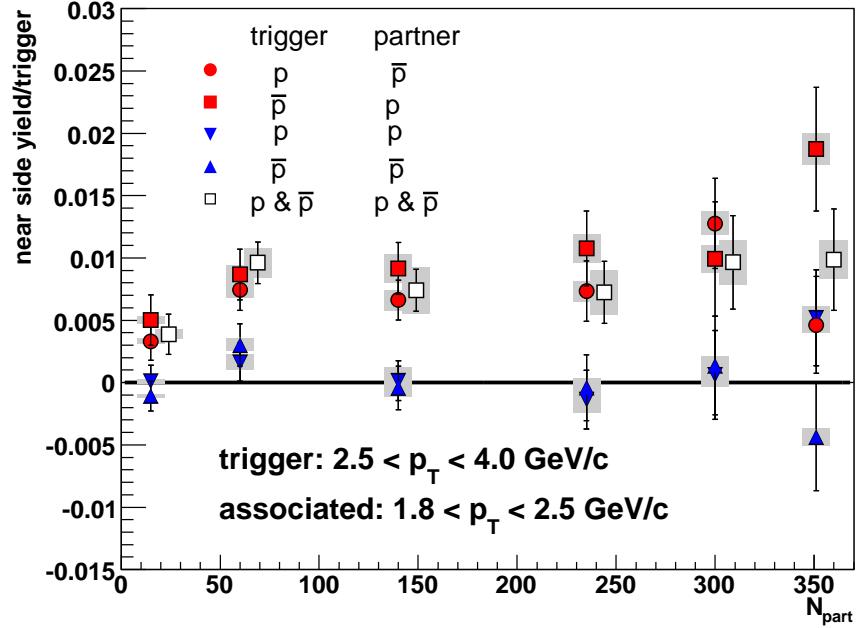}
\caption{Near side conditional yields per trigger for charge selected (filled
points)
 and charge inclusive (hollow points)  $p$ and $\bar{p}$
correlations.  Triggers have 2.5$<p_T<$4.0~\gevc and associated particles
have 1.8$<p_T<$2.5~\gevc.  Error bars are statistical
errors and shaded boxes show systematic errors.  There is a
11.4\% (8.9\%) normalization error on baryon ($p$,$\bar{p}$) associated
particle points.}
\label{ppbar}
\end{figure}

 Since baryon number is
a conserved quantity, measurement of the charge
dependence of correlations between two baryons can be a sensitive probe of differences
in the baryon production mechanism and possibly the jet fragmentation
process.
Fig. \ref{ppbar} shows $J(\Delta\phi)$ integrated for $\Delta\phi<0.94$~rad,
i.e the yield of associated particles per trigger as a function of
the number of participating nucleons, $N_{part}$ in the same \mom range
as Fig.~\ref{fig7}. 
The \dphi region covers where two correlated  particles are expected to come from 
the fragmentation of the same jet.
Again, the background has been subtracted with the absolute subtraction
method.
 Both particles are identified
as $p$ or $\bar{p}$ and different sets of points show
the different charge combinations  with triggers
in the \mom region of the baryon excess.  The charge inclusive baryon-baryon yield
(hollow squares) is flat with $N_{part}$,
 except for a smaller yield in the most peripheral collisions.
Same sign pairs, $p$-$p$ and 
$\bar{p}$-$\bar{p}$ (triangles), show no yield 
and opposite sign pairs (filled circles and squares)
are consistent with the charge independent yields.
No significant difference is seen between $p$ and $\bar{p}$
triggers.

\subsection{Associated Particle Ratios}

\begin{figure}
\includegraphics[width=\textwidth]{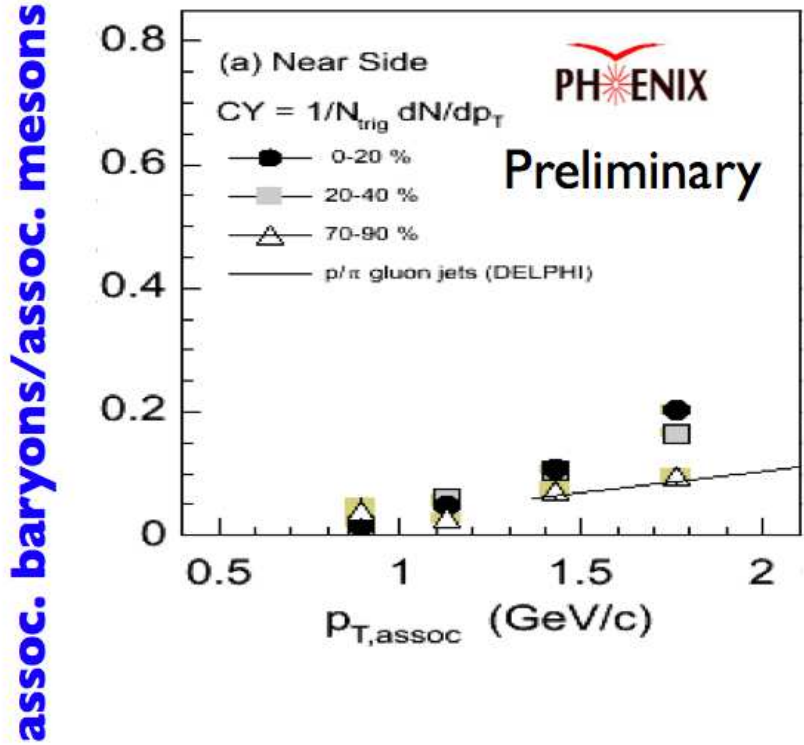}
\caption{ Ratio of near-side associated baryons to mesons as
a function of \mom for three centrality selections.  Trigger
particles are charged hadrons with 2.5$<p_T<$4.0~GeV/$c$. } 
\label{near_ratio}
\end{figure}

\begin{figure}
\includegraphics[width=\textwidth]{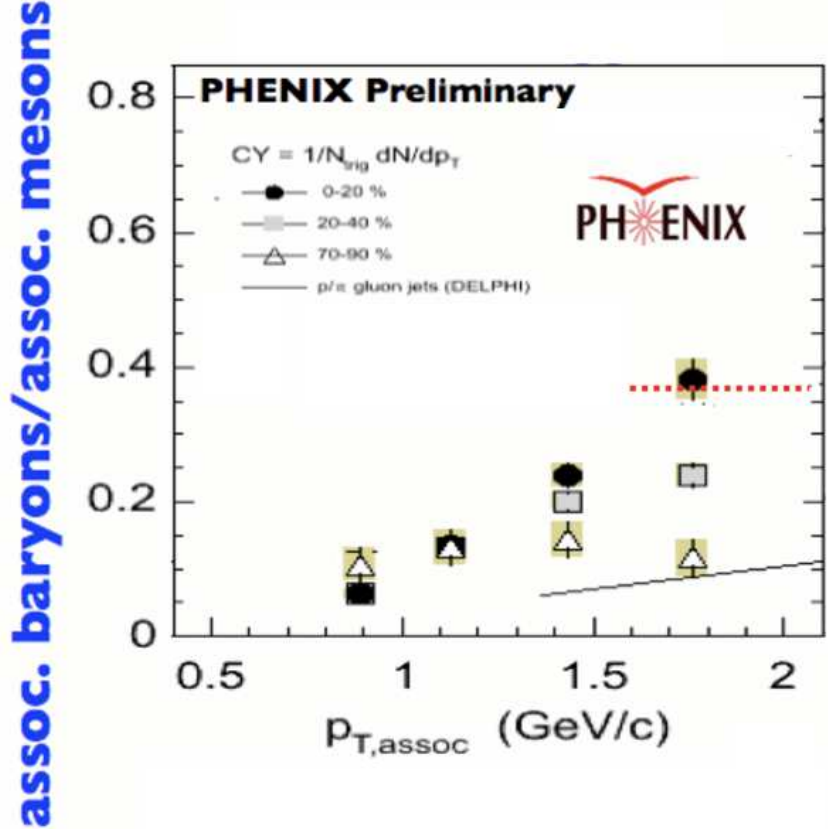}
\caption{ Ratio of away-side associated baryons to mesons as
a function of \mom for three centrality selections.  Trigger
particles are charged hadrons with 2.5$<p_T<$4.0~GeV/$c$. } 
\label{away_ratio}
\end{figure}

Baryons at all centralities are associated with
non-zero jet-like conditional yields, so it is useful to
study whether the particle mixture of jet fragments
changes with centrality. 
Here the triggers are charged hadrons with 2.5$<p_T<$4.0~GeV/$c$,
the same range as in Section~\ref{sectidtrig}.
Due to the baryon excess in central collisions,
the fraction of baryons in
the trigger sample increases with centrality.  The
combinatorial background has been subtracted with the ZYAM
assumption and $J(\Delta\phi)$ has been integrated over
$\Delta\phi$ less than the minimum in $J(\Delta\phi)$.
Fig.~\ref{near_ratio} shows the ratio of associated
baryons to mesons as a function of $p_{T,assoc}$ for
the near-side jet in three centrality classes.  At
low $p_T$ the ratio is small and there is no significant
centrality dependence.  At $p_T>$1.5~\gevc the ratio increases
with increasing centrality.

Triggering on an intermediate \mom particle is expected to bias the
near-side jet toward small medium path lengths.  If so, 
the away-side, $\Delta\phi\approx\pi$, typically sees a long medium path length and
could be sensitive to medium modifications to the jet
fragmentation process, hence we measure the particle composition
of the away-side jet.
Fig. \ref{away_ratio} shows the ratio of associated baryons to 
mesons ($\pi^{\pm}$,$K^{\pm}$) with charged hadron triggers,
2.5$<p_T<$4.0~GeV/$c$, as a function of the associated particle $p_T$
integrated over \dphi from $\pi$ to the minimum of $J(\Delta\phi)$.
 In peripheral collisions (triangles) the
 ratio of associated baryons to mesons on the away-side is
approximately flat with $p_T$.
In central collisions (circles) this ratio increases significantly
with the associated particle $p_T$, suggesting that
the away-side jet fragmentation is increasingly baryon rich 
at intermediate $p_T$.  At the highest associated particle
\mom shown in central collisions the ratio of associated baryons to 
mesons is consistent with the value observed in the single particles at
the same \mom and centrality selections~\cite{ppg026}.

\section{Conclusions}
The extraction of jet-like correlations at intermediate \mom
is difficult because of the large combinatorial background.  
PHENIX has developed robust methods to reliably subtract
this background.  We have described in detail the absolute subtraction
method which makes no assumptions about the jet correlation shape.

We have presented some recent results of identified particle 
jet correlations from PHENIX.  In the same \mom range that 
an excess of baryons has been observed in single particle yields
we have observed modification to the jet structure of
two particle correlations involving baryons in central 
collisions.  These results suggest that, at least some of, the
baryon excess is connected to jet fragmentation in central
Au+Au collisions being modified compared to vacuum fragmentation.
The yield of associated particles per trigger increases with centrality
and the fraction of associated particles per hadron trigger
that are baryons increases.
These observations, along with the quark number scaling observed in elliptic
flow measurements in the same \mom range could indicate that 
particle production at intermediate \mom is a novel interplay
of hard and soft physics.  A full understanding of this
phenomenology will require models which are able to 
simultaneously explain single particle yields,
elliptic flow and jet correlations.  

\bibliographystyle{hunsrt.bst}
\bibliography{sickles_hangzhou}

\end{document}